\documentclass[aps,prd,nofootinbib,showkeys,floatfix]{revtex4}

\usepackage{amsmath}
\usepackage{amssymb}
\usepackage{graphicx}
\usepackage{rotating}
\usepackage{bbm}  
\usepackage{color} 
\usepackage{multirow} 
\usepackage[T1]{fontenc} 
\usepackage{slashed}

\newcommand{\be}{\begin{equation}}
\newcommand{\ee}{\end{equation}}
\newcommand{\ba}{\begin{eqnarray}}
\newcommand{\ea}{\end{eqnarray}}
\newcommand{\no}{\nonumber\\}

\newcommand{\AddrAHEP}{
  {\it AHEP Group, Instituto de F\'{\i}sica Corpuscular --
    C.S.I.C./Universitat de Val{\`e}ncia \\
    Edificio de Institutos de Paterna, Apartado 22085,
  E--46071 Val{\`e}ncia, Spain}}

\newcommand{\AddrLisb}{%
Technical University of Lisbon, CFTP \\
Instituto Superior T\'ecnico, 1049-001 Lisboa, Portugal
}

\newcommand{\AddWurz}{ 
 Institut f{\"u}r Theoretische Physik und
  Astrophysik, \\ Universit{\"a}t W{\"u}rzburg, 97074 W{\"u}rzburg,
  Germany}

\def\gsim{\raise0.3ex\hbox{$\;>$\kern-0.75em\raise-1.1ex\hbox{$\sim\;$}}}
\def\lsim{\raise0.3ex\hbox{$\;<$\kern-0.75em\raise-1.1ex\hbox{$\sim\;$}}}

\begin{document}


\begin{flushright}
CFTP/12-008\\
IFIC/12-32  
\end{flushright}

\title{Accidental stability of dark matter}

\author{L. Lavoura} \email{balio@cftp.ist.utl.pt}
\affiliation{\AddrLisb}
\author{S. Morisi} \email{morisi@ific.uv.es}
\affiliation{\AddrAHEP,\AddWurz}
\author{J. W. F. Valle} \email{valle@ific.uv.es}
\affiliation{\AddrAHEP}

\date{\today}
\begin{abstract}
We propose that dark matter is stable
as a consequence of an \emph{accidental}\/ $\mathbbm{Z}_2$
that results from a flavour symmetry group
which is the double-cover group of the symmetry group
of one of the regular geometric solids. 
Although  model-dependent,
the phenomenology resembles that of a generic ``inert Higgs''
dark matter scheme.

\end{abstract}
\maketitle

\section{Introduction}

The Standard Model (SM) has so far provided a remarkably good
description of Nature,
except for its failure to account for neutrino
oscillations~\cite{art:2012,maltoni:2004ei}
and for the growing evidence
for the existence of dark matter~\cite{Bertone2005279}.
Perhaps these two seemingly unrelated problems
constitute two sides of the same coin.
Several attempts have been
made~\cite{Ma:2001dn,babu:2002dz,altarelli:2005yp,Hirsch:2012ym}
to account for the observed pattern of the neutrino mass and mixing parameters
starting from a fundamental flavour symmetry~\cite{Ishimori:2010au}.
Similarly,
many models have been built to explain
dark matter by invoking supersymmetry and imposing unbroken $R$-parity
in an \textit{ad hoc}\/ fashion~\cite{Drees:1992am,jungman:1996df}.
Various papers in which a $\mathbbm{Z}_2$ has been used
to connect the neutrino masses to dark matter
are given in Refs.~\cite{Ma:2006km,Boehm:2006mi,Deshpande:1977rw,LopezHonorez:2010tb,Gu:2008zf,Farzan:2009ji,Ma:2008cu,Ma:2009gu,Farzan:2012sa,Parida:2011wh,Suematsu:2010nd,Kanemura:2011mw}.

There have been recently many attempts to link neutrino properties to
dark matter by having the latter stabilized by a remnant $\mathbbm{Z}_2$
arising from the underlying flavour
symmetry~\cite{Hirsch:2010ru,Meloni:2011cc,Boucenna:2011tj,Meloni:2010sk}.

Here we attempt a different but related approach,
namely,
we provide a class of models---with a flavour symmetry group
which is the double-cover group of the symmetry group
of one of the regular geometric solids---where dark matter is stabilized
as a result of an \textit{accidental}\/ $\mathbbm{Z}_2$.
We present two examples of this class of models,
namely,
we upgrade two representative $A_4$-based
models~\cite{babu:2002dz,altarelli:2005yp}
to endow them with an automatically stable dark-matter particle.

The plan of the paper is as follows.
In Sec.~\ref{sec:some-mathematics} we present the basic mathematics.
In Sec.~\ref{ourp} we describe our proposal.
In Sec.~\ref{model} we provide two realizations of the proposal,
briefly describing the general aspects of dark-matter phenomenology
in Sec.~\ref{sec:dark-matt-stab}.
Relevant group character tables are collected
in appendix~\ref{sec:character-tables}.
Appendix~\ref{sec:mass-matrix} contains
the derivation of the mass matrix of the scalars in the dark-matter sector
of our first model.

\section{Some mathematics}
\label{sec:some-mathematics}

Here we describe the relevant basic mathematics.
Let
\be
H = \vec \pi \cdot \vec \sigma
= \left( \begin{array}{cc}
\pi_3 & \pi_1 - i \pi_2 \\ \pi_1 + i \pi_2 & - \pi_3
\end{array} \right)
\ee
be a Hermitian matrix---the $\pi_j$ ($j = 1, 2, 3$) are real
quantities and the $\sigma_j$ are the Pauli matrices.
Let $M$ be a matrix of $SU(2)$.
Then,
the transformation
\be
H \to M H M^\dagger = \vec \pi^\prime \cdot \vec \sigma
= \left( \begin{array}{cc}
\pi_3^\prime & \pi_1^\prime - i \pi_2^\prime \\
\pi_1^\prime + i \pi_2^\prime & - \pi_3^\prime
\end{array} \right)
\label{hprime}
\ee
is equivalent to a transformation
\be
\left( \begin{array}{c} \pi_1 \\ \pi_2 \\ \pi_3 \end{array} \right)
\to
\left( \begin{array}{c} \pi_1^\prime \\ \pi_2^\prime \\ \pi_3^\prime
\end{array} \right)
= M^\prime
\left( \begin{array}{c} \pi_1 \\ \pi_2 \\ \pi_3 \end{array} \right),
\label{mprime}
\ee
where $M^\prime \in SO(3)$. 
In this way,
each matrix $M^\prime$ of $SO(3)$ may be mapped into two different matrices,
$M$ and $-M$,
of $SU(2)$.\footnote{More precisely,
$SO(3)$ is isomorphic to $SU(2) / C$,
where $C = \left\{ \mathbbm{1}_2, - \mathbbm{1}_2 \right\}$
is the centre of $SU(2)$,
\textit{i.e.}\ the $\mathbbm{Z}_2$ group formed by the matrices of $SU(2)$
which commute with all the matrices of $SU(2)$.}
One says that $SU(2)$ is the \textit{double cover}
of $SO(3)$.\footnote{Notice,
however,
that $SO(3)$ is \emph{not} a subgroup of $SU(2)$.}

One consequence of this fact is that $SU(2)$
has all the irreducible representations (irreps) as $SO(3)$,
\textit{viz.}\ the \textbf{1},
\textbf{3},
\textbf{5},
and so on,\footnote{We adopt in this paper the standard practice
of denoting an irrep by its dimension in boldface style.}
plus some additional irreps of its own,
\textit{viz.}\ the \textbf{2},
\textbf{4},
\textbf{6},
and so on.
Moreover,
if we call the irreps of $SO(3)$ ``vectorial''
and the extra irreps of $SU(2)$ ``spinorial'',
then the product either of two spinorial irreps or of two vectorial irreps
has only vectorial irreps in its Clebsch--Gordan series,
while the product of one spinorial irrep and one vectorial irrep
has only spinorial irreps in the Clebsch--Gordan series;
it all happens as if there were an \emph{accidental} $\mathbbm{Z}_2$ symmetry
under which the spinorial irreps transformed into minus themselves.

Geometrically,
$SO(3)$ may be interpreted as the group of rotations
in three-dimensional space.
It has three remarkable finite discrete subgroups,
which are the symmetry groups of the five regular geometric solids.
Those subgroups are $A_4$,
which has 12 elements and is the symmetry group of the regular tetrahedron,
$S_4$,
which has 24 elements and is the symmetry group
of the cube and of the regular octahedron,
and $A_5$,
which has 60 elements and is the symmetry group
of the regular dodecahedron and of the regular icosahedron.

These three subgroups of $SO(3)$ each have a double cover in $SU(2)$.
We shall adopt the convention of denoting the double cover
of an $SO(3)$ subgroup by the name of that subgroup with a tilde.
The double-cover groups have twice as many elements as the original group:
$\tilde A_4$ has 24 elements,\footnote{It is identified
by the  group-manipulation software GAP as $[24,3]$ and it is named
$SL(2,3)$ in Ref.~\cite{Parattu:2010cy}.}
$\tilde S_4$ has 48 elements,\footnote{Its GAP identifier is $[48,28]$
and it is named $SL(2,3) \rightarrow G \rightarrow C_2$
in Ref.~\cite{Parattu:2010cy}.}
and $\tilde A_5$ has 120 elements.
The double-cover groups are produced by the same trick
performed to obtain $SU(2)$ from $SO(3)$;
one starts from an irrep \textbf{3} of the $SO(3)$ subgroup,
one interprets each of its matrices $M^\prime$
as a transformation in Eq.~(\ref{mprime}),
and one transforms that $M^\prime$ into two $SU(2)$ matrices $M$ and $-M$
via Eq.~(\ref{hprime});
the matrices thus obtained constitute the defining
two-dimensional irrep of the double-cover group.
Remarkably,
the double-cover groups also have vectorial and spinorial irreps;
the vectorial irreps are identical to the irreps of the $SO(3)$ subgroup,
while the spinorial irreps are extra irreps of the double-cover group.

Let us illustrate this firstly with the group $\tilde A_4$.
In its defining irrep $\mathbf{2}_1$,
it is generated by the $SU(2)$ matrices
\be
M_1 = \left( \begin{array}{cc}
i & 0 \\ 0 & -i \end{array} \right)
\ \mathrm{and} \
M_2 = \frac{1}{\sqrt{2}} \left( \begin{array}{cc}
\sigma & \sigma \\ \sigma^3 & \sigma^7
\end{array} \right),
\ \mathrm{where}\ \sigma = e^{i \pi / 4} = \frac{1+i}{\sqrt{2}}.
\ee
These matrices produce,
via the trick in~Eqs.~(\ref{hprime}) and~(\ref{mprime}),
the matrices
\be
M_1^\prime = \left( \begin{array}{ccc}
-1 & 0 & 0 \\ 0 & -1 & 0 \\ 0 & 0 & 1
\end{array} \right)
\ \mathrm{and} \
M_2^\prime = \left( \begin{array}{ccc}
0 & 1 & 0 \\ 0 & 0 & 1 \\ 1 & 0 & 0
\end{array} \right),
\ee
respectively,
which belong to $SO(3)$ and generate the defining irrep $\mathbf{3}$ of $A_4$,
which is also an irrep of $\tilde A_4$.
The group $\tilde A_4$ has seven inequivalent irreps,
four of which---the $\mathbf{3}$ and the $\mathbf{1}_j$---are also
irreps of $A_4$ and are vectorial,
while the extra three irreps $\mathbf{2}_j$ are not irreps of $A_4$
and are spinorial.
The characters of the irreps of $\tilde A_4$ are given in appendix~A.
The $\mathbf{1}_1$ is the trivial irrep,
in which all the elements of $\tilde A_4$ are represented by the number 1.
The vectorial \textit{vs.}\ spinorial character of the irreps
is visible in the multiplication table given in table~I.
\begin{center}
\begin{table}[!ht]
\begin{tabular}{|c||c|c|c|c||c|c|c|}
\hline
$\otimes$ &
$\mathbf{1}_1$ &
$\mathbf{1}_2$ &
$\mathbf{1}_3$ &
$\mathbf{3}$ &
$\mathbf{2}_1$ &
$\mathbf{2}_2$ &
$\mathbf{2}_3$ \\
\hline \hline
$\mathbf{1}_1$ & $\mathbf{1}_1$ & $\mathbf{1}_2$ & $\mathbf{1}_3$
& $\mathbf{3}$ & $\mathbf{2}_1$ & $\mathbf{2}_2$ & $\mathbf{2}_3$ \\
\hline
$\mathbf{1}_2$ &                & $\mathbf{1}_3$ & $\mathbf{1}_1$
& $\mathbf{3}$ & $\mathbf{2}_2$ & $\mathbf{2}_3$ & $\mathbf{2}_1$ \\
\hline
$\mathbf{1}_3$ &                &                & $\mathbf{1}_2$
& $\mathbf{3}$ & $\mathbf{2}_3$ & $\mathbf{2}_1$ & $\mathbf{2}_2$ \\
\hline
$\mathbf{3}$ & & &
& $\mathbf{3}, \mathbf{3}, \mathbf{1}_1, \mathbf{1}_2, \mathbf{1}_3$
& $\mathbf{2}_1,\mathbf{2}_2, \mathbf{2}_3$
& $\mathbf{2}_1, \mathbf{2}_2, \mathbf{2}_3$
& $\mathbf{2}_1, \mathbf{2}_2, \mathbf{2}_3$ \\
\hline
\hline
$\mathbf{2}_1$
& & & &
& $\mathbf{3}, \mathbf{1}_1$
& $\mathbf{3}, \mathbf{1}_2$
& $\mathbf{3}, \mathbf{1}_3$
\\
\hline
$\mathbf{2}_2$ & & & & & &
$\mathbf{3}, \mathbf{1}_3$ &
$\mathbf{3}, \mathbf{1}_1$
\\
\hline
$\mathbf{2}_3$ & & & & & & &
$\mathbf{3}, \mathbf{1}_2$
\\
\hline
\end{tabular}
\caption{Vectorial \textit{vs.}\ spinorial character 
of the irreps of $\tilde{A}_4$.}
\end{table}
\end{center}
One sees that spinorial irreps are exclusively obtained from the
product of one vectorial and one spinorial irrep.

The same features apply to the double-cover groups of $S_4$ and $A_5$,
the character tables of which are also given in appendix~A.
Note that the first five irreps of $\tilde S_4$ are vectorial
and the latter three are spinorial.
Similarly,
the first five irreps of $\tilde A_5$ are vectorial
and are also irreps of $A_5$,
while the last four irreps of $\tilde A_5$ are spinorial.

Notice that $\tilde S_4$ has three inequivalent doublet irreps,
one of which
(the \textbf{2}$_\mathrm{V}$)
is vectorial while the other two
(the $\mathbf{2}_1$ and the $\mathbf{2}_2$)
are spinorial.
Similarly,
$\tilde A_5$ has two inequivalent quadruplet irreps,
one of which is vectorial
(and is also an irrep of $A_5$)
while the other one is spinorial.

As stressed above,
the double-cover groups $\tilde A_4$,
$\tilde S_4$,
and $\tilde A_5$ are subgroups of $SU(2)$.
The branching rules for the various irreps of $SU(2)$
in irreps of its subgroups are given in table~II.
\begin{center}
\begin{table}[!ht]
\begin{tabular}{|c|c|c|c|}
\hline
$SU(2)$ & $\tilde A_4$ & $\tilde S_4$ & $\tilde A_5$ \\
\hline
$\mathbf{1}$ & $\mathbf{1}_1$ & $\mathbf{1}_1$ & $\mathbf{1}$ \\
$\mathbf{2}$ & $\mathbf{2}_1$ & $\mathbf{2}_1$ & $\mathbf{2}_1$ \\
$\mathbf{3}$ & $\mathbf{3}$ & $\mathbf{3}_1$ & $\mathbf{3}_1$ \\
$\mathbf{4}$ & $\mathbf{2}_2, \mathbf{2}_3$ & $\mathbf{4}$ &
$\mathbf{4}_\mathrm{S}$ \\
$\mathbf{5}$ & $\mathbf{3}, \mathbf{1}_2, \mathbf{1}_3$ &
$\mathbf{3}_2, \mathbf{2}_\mathrm{V}$ & $\mathbf{5}$ \\
$\mathbf{6}$ & $\mathbf{2}_1, \mathbf{2}_2, \mathbf{2}_3$ &
$\mathbf{4}, \mathbf{2}_2$ & $\mathbf{6}$ \\
$\mathbf{7}$ & $\mathbf{3}, \mathbf{3}, \mathbf{1}_1$ &
$\mathbf{3}_1, \mathbf{3}_2, \mathbf{1}_2$ &
$\mathbf{4}_\mathrm{V}, \mathbf{3}_2$ \\
$\mathbf{8}$ & $\mathbf{2}_1, \mathbf{2}_1, \mathbf{2}_2, \mathbf{2}_3$ &
$\mathbf{4}, \mathbf{2}_1, \mathbf{2}_2$ &
$\mathbf{6}, \mathbf{2}_2$ \\
\hline
\end{tabular}\, .
\caption{Branching rules 
for the lowest-dimensional irreps of $SU(2)$.}
\end{table}
\end{center}
One sees that the \textbf{2}$_\mathrm{V}$ of $\tilde S_4$
and the \textbf{4}$_\mathrm{V}$ of $\tilde A_5$
are vectorial---they appear in the branching
of the vectorial irreps \textbf{5} and \textbf{7},
respectively,
of $SU(2)$---in spite of having dimensions
that one associates in the case of $SU(2)$ to spinorial irreps.

\section{Our proposal}
\label{ourp}

The group $A_4$ has been used as horizontal-symmetry group for the
leptonic sector in countless models and papers during the last
decade~\cite{Altarelli:2010gt,Ishimori:2010au,Hirsch:2012ym}.  
It has been used to account for the predictions $\theta_{23}=\pi/4$ and
$\theta_{13}=0$~\cite{babu:2002dz}
as well as to explain the full tri-bimaximal mixing (TBM),
namely the fact that the lepton mixing matrix $U$ is rather close to
\be
U_\mathrm{TBM} = \left( \begin{array}{ccc}
\sqrt{2/3} & \sqrt{1/3} & 0 \\
- \sqrt{1/6} & \sqrt{1/3} & \sqrt{1/2} \\
- \sqrt{1/6} & \sqrt{1/3} & - \sqrt{1/2}
\end{array} \right).
\ee
We can take $U_\mathrm{TBM}$ as a first-order approximation
to the true lepton mixing matrix:
$U\approx U_\mathrm{TBM}$.\footnote{Even in a model
in which $U = U_\mathrm{TBM}$,
this prediction is in general approximate,
since it may only hold at a high-energy scale and it will then be corrected
by the renormalization-group evolution~\cite{antusch:2005gp}
down to low-energy scale.
It may also be corrected by other effects,
for instance a non-diagonal charged-lepton mass matrix.}

The group $A_5$ has also been used as a flavour group for the leptonic sector,
namely in a model~\cite{Everett:2008et}
that predicts $\cos{\theta_{12}} = \varphi$,
where $\theta_{12}$ is the solar-neutrino mixing angle
and $\varphi = \left. \left( 1 + \sqrt{5} \right) \right/ 2$
is the so-called ``golden ratio''~\cite{Kajiyama:2007gx}.
Finally,
the flavour group $S_4$ has also been used
in a few papers~\cite{Lam:2008rs,Bazzocchi:2008ej,Altarelli:2010gt,ivo}.

Our proposal consists in the following.
In any of the flavour models using either an $A_4$,
$S_4$,
or $A_5$ flavour-symmetry group,
one may use instead their double covers $\tilde A_4$,
$\tilde S_4$,
and $\tilde A_5$,
respectively.
This is so because
the vectorial irreps and their respective Clebsch--Gordan series
and coefficients are identical for any group and its double cover.
When one does that,
one obtains a model in which all the `matter',
\textit{i.e.}\ all the fermion and scalar fields,
are in vectorial irreps.
We propose
\emph{to add to any such model some `dark matter'
in spinorial irreps of the flavour group},
\textit{viz.}\ of $\tilde A_4$,
$\tilde S_4$,
or $\tilde A_5$.

It is furthermore crucial that no field of the `dark matter' sector
acquires a vacuum expectation value (VEV).
Indeed,
let $H$ denote a generic field in the `matter' sector
and $\eta$ a generic field in the `dark matter' sector.
Since $\eta$ has a spinorial character---even if it is
an integer-spin field!---under the flavour group,
it will only have $\eta \eta$ and $\eta \eta \eta \eta$ self-interactions,
plus $\eta \eta H$ and $\eta \eta H H$ interactions with the `matter' sector.
The latter interactions,
however,
cannot cause the lightest $\eta$ field to decay,
they can only cause it to co-annihilate.
It follows that the lightest $\eta$ field is stable,
and therefore,
if electrically neutral,
it constitutes a potentially viable dark-matter candidate.
The vectorial \textit{vs.}\ spinorial character of the various irreps
of the flavour group effectively acts as an
(accidental)
$\mathbbm{Z}_2$ symmetry
preventing $\eta H H$ and $\eta H H H$ couplings,
which would cause $\eta$ to decay into matter.
It is also crucial that no $\eta$ field acquires a VEV
$\left\langle \eta \right\rangle_0$,
lest the $\eta \eta H H$ interaction
produces a $\left\langle \eta \right\rangle_0 \eta H H$ interaction
which would cause $\eta$ to decay.

Notice that the fact that
eventually the whole flavour symmetry group ends up being spontaneously broken
is immaterial for the above reasoning.
Indeed,
the one thing that matters is that the flavour symmetry group leads to
\emph{an accidental $\mathbbm{Z}_2$ symmetry which remains unbroken}
and prevents the appearance
in the Lagrangian---even upon renormalization---of terms of the form
$\eta H H$ or $\eta H H H$,
which would cause the fields $\eta$ in the `dark matter' sector
to decay into fields $H$ of the `matter' sector.

\section{Model building} 
\label{model}

We consider in this section,
as explicit examples,
two models based on $A_4$,
which we extend to $\tilde A_4$
in order to include inert dark matter ``accidentally'' stabilized,
as described in the previous section.
In this section we shall use the more usual notation $T'$
to denote the double-cover group $\tilde A_4$ of $A_4$.

\subsection{Model 1} 
\label{sec:model-1}

Consider the model defined by table~III.
\begin{table}[ht!]
\begin{center}
\begin{tabular}{|c||c|c|c||c|c|c|c|c|}
\hline
$ $ & $L$ & $l_R$ & $\nu_R$ & $H$ & $H_T$ & $h$  & $\phi$ & $\eta$ \\
\hline
$SU(2)$ & $\mathbf{2}$ & $\mathbf{1}$ & $\mathbf{1}$ &
$\mathbf{2}$ & $\mathbf{2}$ & $\mathbf{2}$ &
$\mathbf{1}$ & $\mathbf{2}$ \\
\hline
$T'$ & $\mathbf{3}$ & $\mathbf{3}$ & $\mathbf{3}$ &
$\mathbf{1}_1$ & $\mathbf{3}$ & $\mathbf{1}_1$ &
$\mathbf{3}$ & $\mathbf{2}_1$ \\
\hline
$\mathbbm{Z}_2$ & $+$ & $+$ & $-$ & $+$ & $+$ & $-$ & $+$ & +  \\
\hline
\end{tabular}\caption{Matter assignment of model 1.}\label{tabm1}
\end{center}
\end{table}
This model is a generalization
of the model in Ref.~\cite{altarelli:2005yp}.
In table~III,
$SU(2)$ is the Standard Model (SM) gauge group
and there is a flavour symmetry $T'$
and an additional $\mathbbm{Z}_2$ symmetry.
Note that that additional $\mathbbm{Z}_2$
is \emph{not} the accidental $\mathbbm{Z}_2$
that stabilizes the dark matter;
it has been introduced only to obtain the TBM limit.

We observe that the scalar field $\eta$
is the only one that has spinorial character under $T'$,
namely,
it is a doublet of $T'$ and therefore
it cannot interact directly with the SM fermions through Yukawa couplings. 
It can couple to SM particles only through the ``Higgs portal'',
namely via terms like $\eta^\dagger \eta H^\dagger H$
or $\eta^\dagger \eta H_T^\dagger H_T$,
and so on. 
The neutral component of $\eta$ is a good dark matter candidate
since it can be produced in the early universe through the Higgs portal
and since its spinorial character ensures its stability,
as described in Sec.~\ref{ourp}.

The Lagrangian invariant under the SM gauge group
and under the $T'\times \mathbbm{Z}_2$ flavour symmetry is
\begin{eqnarray}
\mathcal{L} &=&
y_H \left( \overline{L} l_R \right)_{\mathbf{1}_1} H
+ y_s \left( \overline{L} l_R \right)_{\mathbf{3}_\mathrm{s}} H_T
+ y_a \left( \overline{L} l_R \right)_{\mathbf{3}_\mathrm{a}} H_T
\no & &
+ y_h \left( \overline{L} \nu_R \right)_{\mathbf{1}_1} h
+ m \left( \nu_R^T C^{-1} \nu_R \right)_{\mathbf{1}_1}
+ y_\nu \left( \nu_R^T C^{-1} \nu_R \right)_{\mathbf{3}_\mathrm{s}} \phi
+ \mathrm{H.c.}
\end{eqnarray}
We assume that the scalar $T'$ triplets have VEVs aligned along the
directions
\begin{equation}
\langle H_T \rangle_0 = v_T \left( 1, 1, 1 \right),
\quad 
\langle \phi \rangle_0 = v \left( 1, 0, 0 \right). 
\end{equation}
Then the charged-lepton mass matrix is given by
\begin{equation}
M_\ell = \left( \begin{array}{ccc}
y_H v_H & \left( y_s+y_a \right) v_T & \left( y_s-y_a \right) v_T \\
\left( y_s-y_a \right) v_T & y_H v_H & \left( y_s+y_a \right) v_T \\
\left( y_s+y_a \right) v_T & \left( y_s-y_a \right) v_T& y_H v_H \\
\end{array} \right),
\end{equation}
which is of the form
\[
\left( \begin{array}{ccc}
\alpha & \beta & \gamma \\
\gamma & \alpha & \beta \\
\beta & \gamma & \alpha
\end{array} \right)
\]
and is diagonalized as
\begin{equation}
U_\omega^\dag M_\ell U_\omega =
\left(
\begin{array}{ccc}
\alpha+\beta+\gamma &0&0\\
0&\alpha+\omega\beta+\omega^2\gamma &0\\
0&0&\alpha+\omega^2 \beta+\omega \gamma 
\end{array}
\right),
\end{equation}
where
\begin{equation}
U_\omega = \frac{1}{\sqrt 3} \left( \begin{array} {c@{\quad}c@{\quad}c}
1 & 1 & 1 \\
1 & \omega & \omega^2 \\
1 & \omega^2 & \omega
\end{array} \right).
\end{equation}
We note that in order to have $m_e\ll m_\mu \ll m_\tau$
a fine-tuning among $\alpha$,
$\beta$,
and $\gamma$ is required:
we need  $\alpha \approx \omega^2 \beta \approx \omega \gamma \sim m_\tau / 3$.
Assuming $v_H \sim 100$ GeV,
the Yukawa couplings must be of order $10^{-2}$
so as to give $y_H v_H \sim m_\tau$.
Thus,
assuming $y_{a,s} \sim \mathcal{O}(1)$,
$v_T$ must be $\mathcal{O}(\mathrm{GeV})$
to give $y_{a,s}v_T\sim m_\tau$.

In the model there are five Higgs doublets---$H$,
three $H_T$,
and $h$---that acquire VEVs,
respectively $v_H$,
$v_T$,
and $v_h$.
These are the VEVs that contribute to the masses of the gauge bosons
$W^\pm$ and $Z^0$.
However,
we assume $v_T, v_h \ll v_H$,
and then $H$ is,
to a good approximation,
the SM Higgs doublet.

We assume that no component of $\eta$ acquires a VEV.
We also assume that the lightest component of $\eta$ is neutral;
that is our dark matter candidate.
The differences among the squared masses
of the various neutral components of $\eta$
are almost of order $\mathcal{O} \left( v_H v_T \right)$,
and then coannihilation is not too strong.
In appendix~\ref{sec:mass-matrix} the form of the mass matrix
of the neutral components of $\eta$ is explicitly computed.

The Dirac neutrino mass matrix is proportional to the unit matrix:
\begin{equation}
M_D \propto \mathcal{I},
\end{equation}
while the right-handed-neutrino Majorana mass matrix is
\begin{equation}
M_R = \left( \begin{array}{ccc}
m & 0 & 0 \\
0 & m & y_\nu v \\
0 & y_\nu v & m
\end{array} \right).
\end{equation}
The light-neutrino Majorana mass matrix arises from the type-I seesaw
mechanism~\cite{schechter:1980gr},
\be
M_\nu = - M_D M_R^{-1} M_D^T \propto M_R^{-1}.
\ee
In the basis where the charged-lepton mass matrix is diagonal,
$\hat{M}_\nu$ is diagonalized by $U_\mathrm{TBM}$
and the mass eigenvalues satisfy
the sum rule~\cite{Barry:2010yk,Dorame:2011eb}
\be
\frac{2}{m_2} = \frac{1}{m_1} + \frac{1}{m_3},
\ee
where the eigenvalues $m_j$ should be understood as being complex,
\textit{viz.}\ the neutrino masses are the $|m_j|$.

In our model we should assign quarks to invariants of $T'$.
In other words,
quarks are flavour-blind and couple only to the SM Higgs doublet $H$,
which is also $T'$-invariant.\footnote{If,
however,
we want to embed the model in a Grand Unified Theory (GUT),
then a possibility is to assign the quarks
to triplets of $T'$ just like the charged leptons.
Then we can in principle implement
either an $SU(5)$ or an $SO(10)$ GUT framework.}
With the above matter assignment,
one may show that the quark mixing matrix
is predicted to be the identity matrix,
which is a good first approximation.
In order to generate both the Cabibbo angle
and the reactor-neutrino mixing angle
we must extend the model in some way.
Note that the data from the recent reactor experiments
Double Chooz~\cite{Abe:2011fz},
Daya Bay~\cite{An:2012eh},
and RENO~\cite{Ahn:2012nd}
seem to indicate that the reactor angle and the Cabibbo angle
are of the same order of magnitude.
Since minimal $SU(5)$ has $M_\ell = M_d^T$,
one may have deviations for lepton mixing
(through the charged-lepton mass matrix $M_\ell$)
and for quark mixing
(through the down-type-quark mass matrix $M_d$)
of the same order --- see for instance
Refs.~\cite{Antusch:2011qg,Antusch:2012fb}.
Another possibility is by assuming
two extra flavon fields $\phi'\sim \mathbf{1}_2$
and $\phi''\sim \mathbf{1}_3$
as in Ref.~\cite{Ma:2011yi},
where it is shown that this leads to a sufficiently large reactor angle.

\subsection{Model 2} 
\label{sec:model-2}

Another $A_4$ model that may be extended to $T'$
in order to accommodate naturally stable inert dark matter
is the model proposed in Refs.~\cite{babu:2002dz,hirsch:2003dr}.
It is described by the field representation content
(in a supersymmetric notation)
in table~IV.
\begin{table}[ht]
\begin{center}
\begin{tabular}{|c||c|c|c|c|c||c|c|}
\hline
 & $\hat Q$ & $\hat L$ &
$ \hat u^c_1, ~\hat d^c_1, ~\hat e^c_1$ &
$ \hat u^c_2, ~\hat d^c_2, ~\hat e^c_2$ &
$ \hat u^c_3, ~\hat d^c_3, ~\hat e^c_3$ &
$\hat \phi_{1,2}$ & $\hat \eta_{1,2}$\\
\hline
$SU(2)$ & $\mathbf{2}$ & $\mathbf{2}$ &
$\mathbf{1}$ & $\mathbf{1}$ & $\mathbf{1}$ & $\mathbf{2}$ & $\mathbf{2}$ \\
\hline
$T'$ & $\mathbf{3}$ & $\mathbf{3}$ & $\mathbf{1}_1$ &$\mathbf{1}_2$ &
$\mathbf{1}_3$& $\mathbf{1}_1$ & $\mathbf{2}_1$ \\
\hline
$\mathbbm{Z}_3$ & $1$ & $1$ & $\omega$ &$\omega$ & $\omega$& $1$ & 1\\
\hline
\end{tabular}\caption{Matter assignment of model~2.}
\end{center}
\end{table}
Once again,
the model has a discrete Abelian symmetry,
in this case $\mathbbm{Z}_3$,
which has nothing to do with the accidental symmetry
that stabilizes dark matter.
Notice that the only field that has spinorial character under $T'$ is $\eta$,
which is a doublet of $T'$ and for that reason has no Yukawa couplings
to the SM fermions.
The stability of $\eta$ is ensured
by its spinorial character under $T'$
and the lightest neutral component of $\eta$
is a dark matter candidate.\footnote{This model
may already possess,
besides $\eta$,
the usual supersymmetric dark matter candidates.}

In addition to the fields in table~IV,
the model of Refs.~\cite{babu:2002dz,hirsch:2003dr}
contains the heavy quark,
heavy lepton,
and Higgs superfields in table~V,
\begin{table}[ht]
\begin{center}
\begin{tabular}{|c|c|c|c|c|c|c|c|c|}
\hline
&$\hat U$&$\hat U^c$ &$\hat D$ &$\hat D^c$&
$\hat E$&$\hat E^c$ &$\hat N^c$ &$\hat \chi$\\
\hline
$A_4$ & $\mathbf{3}$ & $\mathbf{3}$ & $\mathbf{3}$ &
$\mathbf{3}$ & $\mathbf{3}$& $\mathbf{3}$& $\mathbf{3}$& $\mathbf{3}$\\
\hline
$\mathbbm{Z}_3$ & $1$ & $1$ & $1$& $1$&$1$& $1$& $1$&$\omega^2$\\
\hline
\end{tabular}
\caption{Extra matter assignment of the model 2.}
\end{center}
\end{table}
which are all gauge-$SU(2)$ singlets.

The superpotential is given by
\begin{eqnarray}
\hat W &=& M_U \hat U_i \hat U^c_i + f_u \hat Q_i \hat U^c_i \hat \phi_2 +
h^u_{ijk} \hat U_i \hat u^c_j \hat \chi_k
\no & & 
+ M_D \hat D_i \hat D^c_i + f_d \hat Q_i \hat D^c_i \hat \phi_1 +
h^d_{ijk} \hat D_i \hat d^c_j \hat \chi_k
\no & &
+ M_E \hat E_i \hat E^c_i + f_e \hat L_i \hat E^c_i \hat \phi_1 +
h^e_{ijk} \hat E_i \hat e^c_j \hat \chi_k
\no & &
+ M_N \hat N^c_i \hat N^c_i + f_N \hat L_i \hat N^c_i
\hat \phi_2 + \mu \hat \phi_1 \hat \phi_2
\no & &
+ M_\chi \hat \chi_i \hat \chi_i + h_\chi \hat \chi_1
\hat \chi_2 \hat \chi_3.
\end{eqnarray}
One can show that the scalar field $\chi$
may acquire VEV along the $T'$ direction
\be
\langle \chi \rangle_0 \sim \left( 1, 1, 1 \right).
\ee
The charged-lepton masses
are generated after integrating out the heavy $E$ and $E^c$ fields.
The result is given as
\be
{\cal M}_e =
U_\omega \left( \begin{array} {c@{\quad}c@{\quad}c} {h_1^e} & 0 & 0
\\ 0 & {h_2^e} & 0 \\ 0 & 0 & {h_3^e} \end{array} \right)
\frac{\sqrt 3 f_e v_1 u}{M_E}.
\ee
The right-handed-neutrino Majorana mass matrix
is proportional to the identity matrix,
$M_N \propto \mathcal{I}$.
Hence,
after a type-I seesaw,
the light-neutrino Majorana mass matrix is given by
\begin{equation}\label{mnu0}
{\cal M}_\nu = \frac{f_N^2 v_2^2}{M_N}\, U_L^T U_L \propto
\left( \begin{array} {c@{\quad}c@{\quad}c} 1 & 0 & 0 \\ 0 & 0 & 1 \\
0 & 1 & 0 \end{array} \right)
\equiv \lambda.
\end{equation}
This shows that,
at this stage,
neutrinos are degenerate and the atmospheric neutrino mixing angle is maximal.

As we run down to the electroweak scale,
Eq.~(\ref{mnu0}) is corrected by the wavefunction renormalizations of $\nu_e$,
$\nu_\mu$,
and $\nu_\tau$,
as well as by the corresponding vertex renormalizations.
One can then obtain the neutrino squared-mass differences
as well as the solar mixing angle.
In contrast to the previous example,
here the lepton mixing is not predicted to be tri-bimaximal
since the solar angle is left unpredicted.
Given the structure of $\lambda$ at the high scale,
its form at the low scale is fixed to first order as
\be
\lambda_\mathrm{low\, scale} = \left( \begin{array} {c@{\quad}c@{\quad}c}
1 + 2 \delta_{ee} &
\delta_{e \mu} + \delta_{e \tau} &
\delta_{e \mu} + \delta_{e \tau} \\
\delta_{e \mu} + \delta_{e \tau} &
2 \delta_{\mu \tau} &
1 + \delta_{\mu \mu} + \delta_{\tau \tau} \\
\delta_{e \mu} + \delta_{e \tau} &
1 + \delta_{\mu \mu} + \delta_{\tau \tau} &
2 \delta_{\mu \tau}
\end{array} \right),
\label{jvure}
\ee
where we have assumed all the parameters to be real.
The matrix in Eq.~(\ref{jvure}) is obtained by multiplying
the matrix of Eq.~({\ref{mnu0}}) on the left
and on the right by all possible $\nu_i \to \nu_j$ transitions.
The mass matrix in Eq.~(\ref{jvure}) is manifestly $\mu$--$\tau$ symmetric,
yielding maximal atmospheric mixing angle
and a solar mixing angle that can be fitted to the measured value.
As shown in Ref.~\cite{babu:2002dz},
assuming the parameters $\delta_{ij}$ to be complex
a deviation of the reactor angle from zero can be obtained.

\subsection{Dark matter } 
\label{sec:dark-matt-stab}

In model~1 above there are two gauge-$SU(2)$ doublets
placed in the $T'$ doublet $\eta$.
In model~2,
which is supersymmetric,
there are instead four gauge-$SU(2)$ doublets
placed in the $T'$ doublets $\hat \eta_1$ and $\hat \eta_2$.
The models~1 and~2 are just two simple examples realizing our idea in Sec.~III.
In contrast with the inert dark matter scenarios~\cite{barbieri:2006dq},
here the dark matter is stabilized accidentally
in the context of flavour symmetry-based models,
as already mentioned in Sec.~\ref{ourp}.
On the other hand,
just as in the inert dark matter models,
in our models above a ``Higgs portal'' exists,
namely terms of the type $\eta \eta H H$
which connect the dark matter to normal matter,
so dark matter can be produced with a relic abundance
$0.09 \le \Omega h^2\le 0.13$
consistent with the WMAP measurements~\cite{Komatsu:2010fb}.

It is possible to perform a detailed study of the parameter space
of either of the above models,
but that goes way beyond the scope of the present paper.
A calculation has been performed for the discrete dark matter scenario
in Ref.~\cite{Boucenna:2011tj}.
In that scenario,
dark matter belongs to a triplet representation of $\tilde{A}_4$,
instead of the spinorial $\mathbf{2}_1$ representation of $\tilde A_4$
of the models above.
However,
we do not expect substantial differences
from the phenomenological point of view.
By analogy with Refs.~\cite{Djouadi:2011aa,Gustafsson:2012aj,Gil:2012ya},
we expect our dark matter candidate $\eta^0$
to be viable within a mass range of 40 to 80 GeV.

\section{Summary}

In summary,
we have proposed that dark matter is stable
because of an accidental $\mathbbm{Z}_2$ symmetry
which results from a flavour group
which is the double-cover group of the symmetry group
of one of the regular geometric solids.
The phenomenology is similar to that of generic inert dark matter scenarios
with a Higgs portal,
except that it appears here in the framework of
discrete flavour symmetry schemes.

\vskip .5cm

Work supported by the Spanish MEC under grants FPA2008-00319/FPA,
FPA2011-22975,
and MULTIDARK CSD2009-00064
(the Consolider--Ingenio 2010 Programme),
by Prometeo/2009/091 (Generalitat Valenciana),
and by the EU ITN UNILHC PITN-GA-2009-237920. S.M supported also from grant DFG grant WI 2639/4-1.
The work of L.L.\ is supported
by the Portuguese \textit{Funda\c c\~ao para a Ci\^encia e a Tecnologia}
under project PEst-OE/FIS/UI0777/2011.
The work of S.M.\ is supported by a Juan de la Cierva contract.
Both S.M.\ and J.V.\ thank S.~Boucenna and E.~Peinado for discussions.

\begin{appendix}

\setcounter{equation}{0}
\renewcommand{\theequation}{A\arabic{equation}}

\section{Character tables}
\label{sec:character-tables}

Here for completeness we present the character tables of the groups
discussed in the text.
\begin{center}
\begin{table}[ht]
\begin{tabular}{|c||c||c|c|c|c||c|c|c|}
\hline
Class &
$n$ &
$\mathbf{1}_1$ &
$\mathbf{1}_2$ &
$\mathbf{1}_3$ &
$\mathbf{3}$ &
$\mathbf{2}_1$ &
$\mathbf{2}_2$ &
$\mathbf{2}_3$ \\
\hline \hline
$C_1$ & 1 &
1 & 1 & 1 & 3 & 2 & 2 & 2 \\
\hline
$C_2$ & 1 &
1 & 1 & 1 & 3 & $-2$ & $-2$ & $-2$ \\
\hline
$C_3$ & 6 &
1 & 1 & 1 & $-1$ & 0 & 0 & 0 \\
\hline
$C_4$ & 4 &
1 & $\omega$ & $\omega^2$ & 0 & 1 & $\omega$ & $\omega^2$ \\
\hline
$C_5$ & 4 &
1 & $\omega^2$ & $\omega$ & 0 & 1 & $\omega^2$ & $\omega$ \\
\hline
$C_6$ & 4 &
1 & $\omega^2$ & $\omega$ & 0 & $-1$ & $-\omega^2$ & $-\omega$ \\
\hline
$C_7$ & 4 &
1 & $\omega$ & $\omega^2$ & 0 & $-1$ & $-\omega$ & $- \omega^2$ \\
\hline
\end{tabular}\caption{Character table of $\tilde{A}_4$.
Here,
$n$ is the number of elements in each class
and $\omega \equiv \exp{\left( 2 i \pi / 3 \right)}
= \left. \left( - 1 + i \sqrt{3} \right) \right/ 2$.}
\end{table}
\end{center}
\begin{center}
\begin{table}[ht]
\begin{tabular}{|c||c||c|c|c|c|c||c|c|c|}
\hline
Class & $n$ &
$\mathbf{1}_1$ &
$\mathbf{1}_2$ &
$\mathbf{2}_\mathrm{V}$ &
$\mathbf{3}_1$ &
$\mathbf{3}_2$ &
$\mathbf{2}_1$ &
$\mathbf{2}_2$ &
$\mathbf{4}$ \\
\hline \hline
$C_1$ & 1 & $1$ & $1$ & $2$ & $3$ & $3$ & $2$ & $2$ & $4$
\\ \hline
$C_2$ & 1 & $1$ & $1$ & $2$ & $3$ & $3$ & $-2$ & $-2$ & $-4$
\\ \hline
$C_3$ & 6 & $1$ & $-1$ & $0$ & $1$ & $-1$ & $\sqrt{2}$ & $-\sqrt{2}$ & $0$
\\ \hline
$C_4$ & 6 & $1$ & $-1$ & $0$ & $1$ & $-1$ & $-\sqrt{2}$ & $\sqrt{2}$ & $0$
\\ \hline
$C_5$ & 6 & $1$ & $1$ & $2$ & $-1$ & $-1$ & $0$ & $0$ & $0$
\\ \hline
$C_6$ & 12 & $1$ & $-1$ & $0$ & $-1$ & $1$ & $0$ & $0$ & $0$
\\ \hline
$C_7$ & 8 & $1$ & $1$ & $-1$ & $0$ & $0$ & $1$ & $1$ & $-1$
\\ \hline
$C_8$ & 8 & $1$ & $1$ & $-1$ & $0$ & $0$ & $-1$ & $-1$ & $1$
\\ \hline
\end{tabular}\caption{Character table of $\tilde{S}_4$.}
\end{table}
\end{center}
\begin{center}
\begin{table}[ht]
\begin{tabular}{|c||c||c|c|c|c|c||c|c|c|c|}
\hline
Class & $n$ &
$\mathbf{1}$ &
$\mathbf{3}_1$ &
$\mathbf{3}_2$ &
$\mathbf{4}_\mathrm{V}$ &
$\mathbf{5}$ &
$\mathbf{2}_1$ &
$\mathbf{2}_2$ &
$\mathbf{4}_\mathrm{S}$ &
$\mathbf{6}$ \\
\hline \hline
$C_1$ & 1 & $1$ & $3$ & $3$ & $4$ & $5$ & $2$ & $2$ & $4$ & $6$
\\ \hline
$C_2$ & 1 & $1$ & $3$ & $3$ & $4$ & $5$ & $-2$ & $-2$ & $-4$ & $-6$
\\ \hline
$C_3$ & 30 & $1$ & $-1$ & $-1$ & $0$ & $1$ & $0$ & $0$ & $0$ & $0$
\\ \hline
$C_4$ & 20 & $1$ & $0$ & $0$ & $1$ & $-1$ & $1$ & $1$ & $-1$ & $0$
\\ \hline
$C_5$ & 20 & $1$ & $0$ & $0$ & $1$ & $-1$ & $-1$ & $-1$ & $1$ & $0$
\\ \hline
$C_6$ & 12 & $1$ & $-b$ & $-d$ & $-1$ & $0$ & $b$ & $d$ & $-1$ & $1$
\\ \hline
$C_7$ & 12 & $1$ & $-b$ & $-d$ & $-1$ & $0$ & $-b$ & $-d$ & $1$ & $-1$
\\ \hline
$C_8$ & 12 & $1$ & $-d$ & $-b$ & $-1$ & $0$ & $d$ & $b$ & $-1$ & $1$
\\ \hline
$C_9$ & 12 & $1$ & $-d$ & $-b$ & $-1$ & $0$ & $-d$ & $-b$ & $1$ & $-1$
\\ \hline
\end{tabular}\caption{Character table of $\tilde{A}_5$.
Here,
$b = {\displaystyle 2 \cos{\frac{4 \pi}{5}}} =
{\displaystyle \frac{- 1 - \sqrt{5}}{2}}$  and
$d= {\displaystyle 2 \cos{\frac{2 \pi}{5}}}
= {\displaystyle \frac{- 1 + \sqrt{5}}{2}}$.
}
\end{table}
\end{center}

\setcounter{equation}{0}
\renewcommand{\theequation}{B\arabic{equation}}

\section{The neutral-scalar squared-mass matrix in model 1}
\label{sec:mass-matrix}

Let $U_1$ and $U_2$ be the generators of $T'$
(the double-covering group of $A_4$).
The irreps of $T'$ may be given thus:
\be
\begin{array}{rcrclcrcl}
\mathbf{1}_k: & &
U_1 &\to& 1, & & U_2 &\to& \omega^{k-1};
\\*[2mm]
\mathbf{2}_k: & &
U_1 &\to& {\displaystyle
\left( \begin{array}{cc} i & 0 \\ 0 & -i \end{array} \right),} & &
U_2 &\to& {\displaystyle
\frac{\omega^{k-1}}{\sqrt{2}} \left( \begin{array}{cc}
\sigma & \sigma \\ \sigma^3 & \sigma^7 \end{array} \right)
};
\\*[5mm]
\mathbf{3}: & &
U_1 &\to& {\displaystyle
\left( \begin{array}{ccc} -1 & 0 & 0 \\ 0 & -1 & 0 \\ 0 & 0 & 1
\end{array} \right),} & &
U_2 &\to& {\displaystyle
\left( \begin{array}{ccc} 0 & 1 & 0 \\ 0 & 0 & 1 \\ 1 & 0 & 0
\end{array} \right)
},
\end{array}
\ee
for $k = 1, 2, 3$,
where $\omega = \exp{\left( i 2 \pi / 3 \right)}$
and $\sigma = \exp{\left( i \pi / 4 \right)}$.
Let $\left( a, b \right)$ be a $\mathbf{2}_1$
and $\left( x, y, z \right)$ be a $\mathbf{3}$ of $T'$.
Then,
\be
\left( \begin{array}{c}
a z + b x - i b y \\ - b z + a x + i a y
\end{array} \right)\, \mathrm{is \ a}\ \mathbf{2}_1, \quad
\left( \begin{array}{c}
a z + \omega^2 b x - i \omega b y \\ - b z + \omega^2 a x + i \omega a y
\end{array} \right)\, \mathrm{is \ a}\ \mathbf{2}_2, \quad
\left( \begin{array}{c}
a z + \omega b x - i \omega^2 b y \\ - b z + \omega a x + i \omega^2 a y
\end{array} \right)\, \mathrm{is \ a}\ \mathbf{2}_3.
\ee
Let $\left( a, b \right)$ be a $\mathbf{2}_p$
and $\left( a', b' \right)$ be a $\mathbf{2}_q$ of $T'$.
Then,
\be
\left( \begin{array}{c}
\omega^{p+q-2} \left( a a' - b b' \right) \\
i \omega^{2 \left( p+q \right) - 2} \left( a a'+ b b' \right) \\
- a b'- b a'
\end{array} \right)\, \mbox{is\ a}\ \mathbf{3}.
\ee

We consider a simplified version of our model~1
by neglecting the scalars $h$ and $\phi$ in table~III.
We then have a six-Higgs-doublet model,
where the Higgs doublets are in a $\mathbf{1}_1$,
a $\mathbf{3}$,
and a $\mathbf{2}_1$ of $T'$,
denoted respectively $H$,
$H_T$,
and $\eta$ in table~III.
Let then
\be
\phi_0:\ \mathbf{1}_1 \ \mathrm{of}\ T', \quad
\left( \begin{array}{c} \eta_1 \\ \eta_2 \end{array} \right):\
\mathbf{2}_1 \ \mathrm{of}\ T', \quad
\left( \begin{array}{c} \phi_1 \\ \phi_2 \\ \phi_3 \end{array} \right):\
\mathbf{3} \ \mathrm{of}\ T'
\ee
be Higgs doublets.
Then,
\be
\phi_0^\dagger:\ \mathbf{1}_1 \ \mathrm{of}\ T', \quad
\left( \begin{array}{c} \eta_2^\dagger \\ - \eta_1^\dagger \end{array} \right):\
\mathbf{2}_1 \ \mathrm{of}\ T', \quad
\left( \begin{array}{c}
\phi_1^\dagger \\ \phi_2^\dagger \\ \phi_3^\dagger \end{array} \right):\
\mathbf{3} \ \mathrm{of}\ T'.
\ee
Making the products of these,
one obtains the following irreps of $T$:
\be
\mathbf{1}_1:\quad
\phi_0^\dagger \phi_0,
\quad
\eta_1^\dagger \eta_1 + \eta_2^\dagger \eta_2,
\quad
\phi_1^\dagger \phi_1 + \phi_2^\dagger \phi_2 + \phi_3^\dagger \phi_3;
\ee
\be
\mathbf{1}_2:\quad
\phi_1^\dagger \phi_1 + \omega^2 \phi_2^\dagger \phi_2
+ \omega \phi_3^\dagger \phi_3;
\ee
\be
\mathbf{1}_3:\quad
\phi_1^\dagger \phi_1 + \omega \phi_2^\dagger \phi_2
+ \omega^2 \phi_3^\dagger \phi_3;
\ee
\be
\mathbf{2}_1:\quad
\left( \begin{array}{c}
\eta_2^\dagger \phi_0 \\ - \eta_1^\dagger \phi_0
\end{array} \right),
\quad
\left( \begin{array}{c}
\phi_0^\dagger \eta_1 \\ \phi_0^\dagger \eta_2
\end{array} \right),
\quad
\left( \begin{array}{c}
\eta_2^\dagger \phi_3 - \eta_1^\dagger \phi_1 + i \eta_1^\dagger \phi_2 \\
\eta_1^\dagger \phi_3 + \eta_2^\dagger \phi_1 + i \eta_2^\dagger \phi_2
\end{array} \right),
\quad
\left( \begin{array}{c}
\phi_3^\dagger \eta_1 + \phi_1^\dagger \eta_2 - i \phi_2^\dagger \eta_2 \\
- \phi_3^\dagger \eta_2 + \phi_1^\dagger \eta_1 + i \phi_2^\dagger \eta_1
\end{array} \right);
\ee
\be
\mathbf{2}_2:\quad
\left( \begin{array}{c}
\eta_2^\dagger \phi_3 - \omega^2 \eta_1^\dagger \phi_1
+ i \omega \eta_1^\dagger \phi_2 \\
\eta_1^\dagger \phi_3 + \omega^2 \eta_2^\dagger \phi_1
+ i \omega \eta_2^\dagger \phi_2
\end{array} \right),
\quad
\left( \begin{array}{c}
\phi_3^\dagger \eta_1 + \omega^2 \phi_1^\dagger \eta_2
- i \omega \phi_2^\dagger \eta_2 \\
- \phi_3^\dagger \eta_2 + \omega^2 \phi_1^\dagger \eta_1
+ i \omega \phi_2^\dagger \eta_1
\end{array} \right);
\ee
\be
\mathbf{2}_3:\quad
\left( \begin{array}{c}
\eta_2^\dagger \phi_3 - \omega \eta_1^\dagger \phi_1
+ i \omega^2 \eta_1^\dagger \phi_2 \\
\eta_1^\dagger \phi_3 + \omega \eta_2^\dagger \phi_1
+ i \omega^2 \eta_2^\dagger \phi_2
\end{array} \right),
\quad
\left( \begin{array}{c}
\phi_3^\dagger \eta_1 + \omega \phi_1^\dagger \eta_2
- i \omega^2 \phi_2^\dagger \eta_2 \\
- \phi_3^\dagger \eta_2 + \omega \phi_1^\dagger \eta_1
+ i \omega^2 \phi_2^\dagger \eta_1
\end{array} \right);
\ee
\be
\mathbf{3}:\quad
\left( \begin{array}{c}
\eta_2^\dagger \eta_1 + \eta_1^\dagger \eta_2 \\
i \omega^2 \left( \eta_2^\dagger \eta_1 - \eta_1^\dagger \eta_2 \right) \\
\eta_1^\dagger \eta_1 - \eta_2^\dagger \eta_2 
\end{array} \right),
\quad
\left( \begin{array}{c}
\phi_2^\dagger \phi_3 \\ \phi_3^\dagger \phi_1 \\ \phi_1^\dagger \phi_2
\end{array} \right),
\quad
\left( \begin{array}{c}
\phi_3^\dagger \phi_2 \\ \phi_1^\dagger \phi_3 \\ \phi_2^\dagger \phi_1
\end{array} \right),
\quad
\left( \begin{array}{c}
\phi_0^\dagger \phi_1 \\ \phi_0^\dagger \phi_2 \\ \phi_0^\dagger \phi_3
\end{array} \right),
\quad
\left( \begin{array}{c}
\phi_1^\dagger \phi_0 \\ \phi_2^\dagger \phi_0 \\ \phi_3^\dagger \phi_0
\end{array} \right).
\ee
Making $\eta_{1,2} \to \eta_{1,2}^0$,
$\phi_0 \to v_H$,
and $\phi_{1,2,3} \to v_T$,
one obtains
\be
\mathbf{1}_1: \quad
\left| v_H \right|^2,
\quad
\left| v_T \right|^2,
\quad
\left| \eta_1^0 \right|^2 + \left| \eta_2^0 \right|^2;
\ee
\be
\mathbf{2}_1: \quad
v_H \left( \begin{array}{c}
{\eta_2^0}^\ast \\ - {\eta_1^0}^\ast
\end{array} \right),
\quad
v_H^\ast \left( \begin{array}{c}
\eta_1^0 \\ \eta_2^0
\end{array} \right),
\quad
v_T \left( \begin{array}{c}
{\eta_2^0}^\ast + \left( i - 1 \right) {\eta_1^0}^\ast \\
{\eta_1^0}^\ast + \left( i + 1 \right) {\eta_2^0}^\ast \\
\end{array} \right),
\quad
v_T^\ast \left( \begin{array}{c}
\eta_1^0 + \left( 1 - i \right) \eta_2^0 \\
- \eta_2^0 + \left( 1 + i \right) \eta_1^0
\end{array} \right);
\label{20}
\ee
\be
\mathbf{2}_2: \quad
v_T \left( \begin{array}{c}
{\eta_2^0}^\ast + \left( i \omega - \omega^2 \right) {\eta_1^0}^\ast \\
{\eta_1^0}^\ast + \left( i \omega + \omega^2 \right) {\eta_2^0}^\ast \\
\end{array} \right),
\quad
v_T^\ast \left( \begin{array}{c}
\eta_1^0 + \left( \omega^2 - i \omega \right) \eta_2^0 \\
- \eta_2^0 + \left( \omega^2 + i \omega \right) \eta_1^0
\end{array} \right);
\label{21}
\ee
\be
\mathbf{2}_3: \quad
v_T \left( \begin{array}{c}
{\eta_2^0}^\ast + \left( i \omega^2 - \omega \right) {\eta_1^0}^\ast \\
{\eta_1^0}^\ast + \left( i \omega^2 + \omega \right) {\eta_2^0}^\ast \\
\end{array} \right),
\quad
v_T^\ast \left( \begin{array}{c}
\eta_1^0 + \left( \omega - i \omega^2 \right) \eta_2^0 \\
- \eta_2^0 + \left( \omega + i \omega^2 \right) \eta_1^0
\end{array} \right);
\label{22}
\ee
\be
\mathbf{3}: \quad
\left( \begin{array}{c}
{\eta_2^0}^\ast \eta_1^0 + {\eta_1^0}^\ast \eta_2^0 \\
i \left( {\eta_2^0}^\ast \eta_1^0 - {\eta_1^0}^\ast \eta_2^0 \right) \\
\left| \eta_1^0 \right|^2 - \left| \eta_2^0 \right|^2
\end{array} \right),
\quad
\left( \begin{array}{c}
\left| v_T \right|^2 \\ \left| v_T \right|^2 \\ \left| v_T \right|^2
\end{array} \right),
\quad
\left( \begin{array}{c}
v_H^\ast v_T \\ v_H^\ast v_T \\ v_H^\ast v_T \end{array} \right),
\quad
\left( \begin{array}{c}
v_H v_T^\ast \\ v_H v_T^\ast \\ v_H v_T^\ast \end{array} \right).
\ee
Therefore,
the quartic terms in the $T'$-invariant scalar potential
yield \emph{only} the following mass terms for $\eta_1^0$ and $\eta_2^0$
when $\phi_0^0$ acquires VEV $v_H$
and $\phi_{1,2,3}^0$ acquire identical VEVs $v_T$:
\ba
& a \left( \left| \eta_1^0 \right|^2 + \left| \eta_2^0 \right|^2 \right), &
\label{mass1} \\
& b \left[ \left| \eta_1^0 \right|^2 - \left| \eta_2^0 \right|^2
+ \left( 1 + i \right) \eta_1^0 {\eta_2^0}^\ast
+ \left( 1 - i \right) {\eta_1^0}^\ast \eta_2^0 \right], &
\label{mass2} \\
& \displaystyle{\frac{c + i d}{2} \left[ 2 \eta_1^0 \eta_2^0
+ \left( - 1 - i \right) \left( \eta_1^0 \right)^2
+ \left( 1 - i \right) \left( \eta_2^0 \right)^2 \right],} &
\label{mass3} \\
& \displaystyle{\frac{c - i d}{2} \left[ 2 {\eta_1^0}^\ast {\eta_2^0}^\ast
+ \left( - 1 + i \right) \left( {\eta_1^0}^\ast \right)^2
+ \left( 1 + i \right) \left( {\eta_2^0}^\ast \right)^2 \right],} &
\label{mass4}
\ea
where $a$,
$b$,
$c$,
and $d$ are real quantities with mass-squared dimension.
We may write the mass terms in equations~(\ref{mass1})--(\ref{mass4})
in the form
\be
\left( \begin{array}{cccc}
\Re{\eta_1^0} & \Re{\eta_2^0} & \Im{\eta_1^0} & \Im{\eta_2^0}
\end{array} \right)
\left( \begin{array}{cccc}
a + b - c + d & b + c & c + d & b - d \\
b + c & a - b + c + d & - b - d & - c + d \\
c + d & - b - d & a + b + c - d & b - c \\
b - d & - c + d & b - c & a - b - c - d
\end{array} \right)
\left( \begin{array}{c}
\Re{\eta_1^0} \\ \Re{\eta_2^0} \\ \Im{\eta_1^0} \\ \Im{\eta_2^0}
\end{array} \right).
\label{mass}
\ee
It is easy to convince oneself that the squared-mass matrix
in Eq.~(\ref{mass}),
even though quite restrictive,
still allows the four neutral components of $\eta$ to be non-degenerate.
In our model,
we should allow the term in Eq.~(\ref{mass1}) to be dominant,
while $b$,
$c$,
and $d$ in Eqs.~(\ref{mass2})--(\ref{mass4})
are $\mathcal{O} \left( v_H v_T \right)$
and provide the non-degeneracy.

\end{appendix}

\bibliographystyle{h-physrev4}

\begin{thebibliography}{10}

\bibitem{art:2012}
A.~McDonald,
talk at the XIV International Workshop on Neutrino Telescopes,
Venice,
March 2011.

\bibitem{maltoni:2004ei}
M.~Maltoni, T.~Schwetz, M.~A. T\'ortola and J.~W.~F. Valle,
New J. Phys. {\bf 6}, 122 (2004) [hep-ph/0405172].

\bibitem{Bertone2005279}
G.~Bertone, D.~Hooper and J.~Silk,
Phys.\ Rept.\ {\bf 405}, 279 (2005).

\bibitem{Ma:2001dn} 
E.~Ma and G.~Rajasekaran,
Phys.\ Rev.\ D {\bf 64}, 113012 (2001) [hep-ph/0106291].

\bibitem{babu:2002dz}
K.~S. Babu, E.~Ma and J.~W.~F. Valle,
Phys. Lett. B {\bf 552}, 207 (2003) [hep-ph/0206292].

\bibitem{altarelli:2005yp}
G.~Altarelli and F.~Feruglio,
Nucl. Phys. B {\bf 720}, 64 (2005) [hep-ph/0504165].

\bibitem{Hirsch:2012ym}
M.~Hirsch \textit{et al.},
1201.5525.

\bibitem{Ishimori:2010au}
H.~Ishimori \textit{et al.},
Prog. Theor. Phys. Suppl. {\bf 183}, 1 (2010) [1003.3552].

\bibitem{Drees:1992am}
M.~Drees and M.~M. Nojiri,
Phys. Rev. D {\bf 47}, 376 (1993) [hep-ph/9207234].

\bibitem{jungman:1996df}
G.~Jungman, M.~Kamionkowski and K.~Griest,
Phys. Rept. {\bf 267}, 195 (1996) [hep-ph/9506380].

\bibitem{Ma:2006km} 
E.~Ma,
Phys.\ Rev.\ D {\bf 73}, 077301 (2006) [hep-ph/0601225].

\bibitem{Boehm:2006mi} 
C.~Boehm, Y.~Farzan, T.~Hambye, S.~Palomares-Ruiz and S.~Pascoli,
Phys.\ Rev.\ D {\bf 77}, 043516 (2008) [hep-ph/0612228].

\bibitem{Deshpande:1977rw} 
N.~G.~Deshpande and E.~Ma,
Phys.\ Rev.\ D {\bf 18}, 2574 (1978)

\bibitem{LopezHonorez:2010tb} 
L.~L\'opez Honorez and C.~E.~Yaguna,
JCAP {\bf 1101}, 002 (2011) [arXiv:1011.1411 [hep-ph]].

\bibitem{Gu:2008zf} 
P.~-H.~Gu and U.~Sarkar,
Phys.\ Rev.\ D {\bf 78}, 073012 (2008) [0807.0270].

\bibitem{Farzan:2009ji} 
Y.~Farzan,
Phys.\ Rev.\ D {\bf 80}, 073009 (2009) [0908.3729].

\bibitem{Ma:2008cu} 
E.~Ma and D.~Suematsu,
Mod.\ Phys.\ Lett.\ A {\bf 24}, 583 (2009) [0809.0942].

\bibitem{Ma:2009gu} 
E.~Ma,
Phys.\ Rev.\ D {\bf 80}, 013013 (2009) [0904.4450].

\bibitem{Farzan:2012sa} 
Y.~Farzan and E.~Ma,
Phys.\ Rev.\ D {\bf 86}, 033007 (2012) [1204.4890].

\bibitem{Parida:2011wh} 
M.~K.~Parida,
Phys.\ Lett.\ B {\bf 704}, 206 (2011) [arXiv:1106.4137 [hep-ph]].

\bibitem{Suematsu:2010nd} 
D.~Suematsu and T.~Toma,
Nucl.\ Phys.\ B {\bf 847}, 567 (2011) [arXiv:1011.2839 [hep-ph]].

\bibitem{Kanemura:2011mw} 
S.~Kanemura, T.~Nabeshima and H.~Sugiyama,
Phys.\ Rev.\ D {\bf 85}, 033004 (2012) [arXiv:1111.0599 [hep-ph]].

\bibitem{Hirsch:2010ru}
M.~Hirsch, S.~Morisi, E.~Peinado and J.W.F.~Valle,
Phys. Rev. D {\bf 82}, 116003 (2010) [1007.0871].

\bibitem{Meloni:2011cc}
D.~Meloni, S.~Morisi and E.~Peinado,
Phys. Lett. B {\bf 703}, 281 (2011) [1104.0178].

\bibitem{Boucenna:2011tj}
M.~Boucenna \textit{et al.},
JHEP {\bf 1105}, 037 (2011) [1101.2874].

\bibitem{Meloni:2010sk}
D.~Meloni, S.~Morisi and E.~Peinado,
Phys. Lett. B {\bf 697}, 339 (2011) [1011.1371].

\bibitem{Parattu:2010cy}
K.~M. Parattu and A.~Wingerter,
Phys. Rev. D {\bf 84}, 013011 (2011) [1012.2842].

\bibitem{Altarelli:2010gt}
G.~Altarelli and F.~Feruglio,
Rev. Mod. Phys. {\bf 82}, 2701 (2010) [1002.0211].

\bibitem{ivo}
I.~de Medeiros Varzielas and L.~Lavoura,
arXiv:1212.3247 [hep-ph].

\bibitem{antusch:2005gp}
S.~Antusch, J.~Kersten, M.~Lindner, M.~Ratz and M.~A. Schmidt,
JHEP {\bf 03}, 024 (2005) [hep-ph/0501272].

\bibitem{Everett:2008et}
L.~L. Everett and A.~J. Stuart,
Phys. Rev. D {\bf 79}, 085005 (2009) [0812.1057].

\bibitem{Kajiyama:2007gx}
Y.~Kajiyama, M.~Raidal and A.~Strumia,
Phys. Rev. D {\bf 76}, 117301 (2007) [0705.4559].

\bibitem{Bazzocchi:2008ej} 
F.~Bazzocchi and S.~Morisi,
Phys.\ Rev.\ D {\bf 80}, 096005 (2009) [arXiv:0811.0345 [hep-ph]].

\bibitem{Lam:2008rs} 
C.~S.~Lam,
Phys.\ Rev.\ Lett.\ {\bf 101}, 121602 (2008) [arXiv:0804.2622 [hep-ph]].

\bibitem{schechter:1980gr}
J.~Schechter and J.~W.~F. Valle,
Phys. Rev. D {\bf 22}, 2227 (1980); 
Phys. Rev. D {\bf 25}, 774 (1982).

\bibitem{Barry:2010yk}
J.~Barry and W.~Rodejohann,
Nucl. Phys. B {\bf 842}, 33 (2011) [1007.5217].

\bibitem{Dorame:2011eb}
L.~Dorame, D.~Meloni, S.~Morisi, E.~Peinado and J.W.F.~Valle,
Nucl.\ Phys.\ B {\bf 861} (2012) 259 [1111.5614].

\bibitem{Abe:2011fz}
DOUBLE-CHOOZ Collaboration, Y.~Abe \textit{et al.},
Phys. Rev. Lett. {\bf 108}, 131801 (2012).

\bibitem{An:2012eh}
DAYA-BAY Collaboration, F.~An \textit{et al.},
Phys. Rev. Lett. {\bf 108}, 171803 (2012) [1203.1669].

\bibitem{Ahn:2012nd}
RENO Collaboration, J.~K.~Ahn \textit{et al.},
Phys.\ Rev.\ Lett.\  {\bf 108} (2012) 191802 [1204.0626].

\bibitem{Antusch:2011qg} 
S.~Antusch and V.~Maurer,
Phys.\ Rev.\ D {\bf 84}, 117301 (2011) [arXiv:1107.3728 [hep-ph]].

\bibitem{Antusch:2012fb} 
S.~Antusch, C.~Gross, V.~Maurer and C.~Sluka,
Nucl.\ Phys.\ B {\bf 866} (2013) 255 [1205.1051].

\bibitem{Ma:2011yi} 
E.~Ma and D.~Wegman,
Phys.\ Rev.\ Lett. {\bf 107}, 061803 (2011) [arXiv:1106.4269 [hep-ph]].

\bibitem{hirsch:2003dr}
M.~Hirsch \textit{et al.},
Phys. Rev. D {\bf 69}, 093006 (2004) [hep-ph/0312265].

\bibitem{barbieri:2006dq}
R.~Barbieri, L.~J. Hall and V.~S. Rychkov,
Phys. Rev. D {\bf 74}, 015007 (2006) [hep-ph/0603188].

\bibitem{Komatsu:2010fb}
WMAP Collaboration, E.~Komatsu \textit{et al.},
Astrophys. J. Suppl. {\bf 192}, 18 (2011) [1001.4538].

\bibitem{Djouadi:2011aa} 
A.~Djouadi, O.~Lebedev, Y.~Mambrini and J.~Quevillon,
Phys.\ Lett.\ B {\bf 709}, 65 (2012) [arXiv:1112.3299 [hep-ph]].

\bibitem{Gustafsson:2012aj} 
M.~Gustafsson, S.~Rydbeck, L.~L\'opez-Honorez and E.~Lundstrom,
Phys.\ Rev.\ D {\bf 86}, 075019 (2012) [arXiv:1206.6316 [hep-ph]].

\bibitem{Gil:2012ya}
G.~Gil, P.~Chankowski and M.~Krawczyk,
Phys.\ Lett.\ B {\bf 717}, 396 (2012) [arXiv:1207.0084 [hep-ph]].

\end{thebibliography}

\end{document}